# Activation of Microwave Fields in a Spin-Torque Nano-Oscillator by Neuronal Action Potentials

J. M. Algarin[1], B. Ramaswamy[2], L. Venuti[3,4], M. E. Swierzbinski[3,4], J. Baker-McKee[7], I. N. Weinberg[7], Y.J. Chen[8], I. N. Krivorotov[8], J. A. Katine[9], J. Herberholz[3,4], R. C. Araneda[4,5], B. Shapiro[2,6] and E. Waks[1,a)]

[1]Institute for Research in Electronics and Applied Physics (IREAP), [2]Fischell Department of Bioengineering, [3]Department of Psychology, [4]Neuroscience and Cognitive Science Program, [5]Department of Biology, [6]Institute for Systems Research (ISR), University of Maryland, College Park, Maryland, 20742, United States.

[7]Weinberg Medical Physics Inc., Bethesda, Maryland, 20852, United States.

[8]Department of Physics and Astronomy, University of California, Irvine, California, 92697, United States

[9]HGST Research Center, San Jose, California, 95135, United States.

**Abstract**

Action potentials are the basic unit of information in the nervous system and their reliable detection and decoding holds the key to understanding how the brain generates complex thought and behavior. Transducing these signals into microwave field oscillations can enable wireless sensors that report on brain activity through magnetic induction. In the present work we demonstrate that action potentials from crayfish lateral giant neuron can trigger microwave oscillations in spin-torque nano-oscillators. These nano-scale devices take as input small currents and convert them to microwave current oscillations that can wirelessly broadcast neuronal activity, opening up the possibility for compact neuro-sensors. We show that action potentials activate microwave oscillations in spin-torque nano-oscillators with an amplitude that follows the action potential signal, demonstrating that the device has both the sensitivity and temporal resolution to respond to action potentials from a single neuron. The activation of magnetic oscillations by action potentials, together with the small footprint and the high frequency tunability, makes these devices promising candidates for high resolution sensing of bioelectric signals from neural tissues. These device attributes may be useful for design of high-throughput bi-directional brain-machine interfaces.

**Introduction**

At its core, the brain is a complex network of neurons connected by synapses. Action potentials are the fundamental units of communication between neurons that form the basic building blocks for thought and behavior [1]–[3]. Detecting these action potentials wirelessly with high spatial and temporal resolution is highly useful to understand how the brain processes information and thought [1], as well as to diagnose and treat neurological diseases [4]–[6].

A number of different techniques exist for wirelessly measuring human brain activity. For example, functional magnetic resonance imaging provides wireless measurements with spatial resolutions on the order of millimeters [7]. However, this technique only measures brain activity indirectly through hemodynamic effects. Furthermore, it does not have the spatial and temporal resolution to isolate single neurons or small clusters and read out individual action potentials [8], [9]. Other methods such as magnetoencephalography provide excellent temporal resolution in the milliseconds range but exhibit very poor spatial resolution [10]. Combining functional magnetic resonance imaging with electroencephalography or magnetoencephalography could potential yield improved temporal

[a)]Author to whom correspondence should be addressed. Electronic mail: edowaks@umd.edu

resolution[11], but spatial resolution still remains in the millimeters range. Currently, the most advanced method for performing highly localized measurements of neuronal action potentials in humans and other primates involves surgical implantation of electrodes to target areas of the brain [12]. Dongjin et al. demonstrated an ultrasonic backscatter system that enables communication with such implanted bioelectronics in the peripheral nervous system [13]. Other techniques like optical methods based on voltage sensitive contrast agents (dyes, quantum dots) and optogenetics have also been demonstrated [14]–[17]. But focused optical beams cannot penetrate the skull or deep tissue, and thus cannot access deep-brain regions [18]. Currently, there is great need in neuroscience for new methods to transduce biological activity to wireless signals that can penetrate through deep tissue.

The application of spintronics to biological sensing remains a relatively unexplored area that has potential to resolve some of the difficult challenges inherent to wireless signal detection. Recent work incorporated giant magnetoresistors into electrode arrays to perform magnetoencephalography [19], but this technique is not wireless. Another compelling spintronic device is the spin-torque nano-oscillator (STNO), which takes as an input small direct currents and converts them to microwave oscillations [20]–[24] that can report wirelessly to a receiver by electromagnetic coupling [25]. The STNO is nanoscale in dimensions and can operate at microwave frequencies ranging from 0.1 – 10 GHz. This property resolves the long-standing challenge inherent to oscillators based on electrical LC circuits that are difficult to scale down to small dimensions. A standard LC circuit of 10 μm dimensions typically exhibits oscillation frequencies exceeding 100 GHz due to limits in achievable values of inductance and capacitance [13]. But these frequencies are incompatible with biological tissues, which become highly absorbing above 5 GHz. Spin-torque oscillators can operate at biologically compatible frequencies while maintaining nanoscale dimensions. Furthermore, they can operate with small input currents, on the order of a micro-amp, which may be sufficiently low to be directly driven by neurons without the need for amplifiers. Finally, the oscillation frequency of the device shifts in the presence of an external magnetic field [22], [26], enabling the precession frequency to encode spatial information by applying a magnetic field gradient, analogous to conventional magnetic resonance imaging. These properties make STNOs promising candidates for detecting weak bioelectric signals with high spatial precision, potentially, up to single cell resolution. But the ability of STNOs to transduce biological signals to microwave fields remains an unexplored area.

Here, we demonstrate that a STNO can transduce a biological signal to microwave field oscillations. We drive the device with action potentials from crayfish neurons. Crayfish possess giant neurons that generate voltages on the order of a few millivolts when measured with extracellular recording electrodes, making them an ideal system to study spintronic devices. We utilized the extracellular voltage produced by the lateral giant neuron to drive the device and observed a clear microwave signal whose temporal envelope accurately reproduced the action potential waveform. This result shows that spintronic devices could potentially serve as nanoscale sensors for bioelectric signals with high spatial resolution and sufficient bandwidth to temporally resolve neuronal action potentials.

The lateral giant escape circuit of crayfish is one of the best understood neuron circuits in the animal kingdom [27], [28]. The key element is a pair of lateral giant neurons. The lateral giants receive mechanosensory inputs in all abdominal segments and produce single action potentials that propagate along the entire ventral nerve cord, the caudal part of the crayfish nervous system, to activate flexor motor neurons [29], [30]. In freely behaving animals this leads to a rapid flexion of the tail and a stereotyped forward "tail-flip" that thrusts the animal away from an attacking predator [31]. The lateral giants are the largest neurons in the ventral nerve cord with axon diameters of up to 200 μm in adult crayfish and can be readily stimulated with extracellular silver wire electrodes both in intact animals and in isolated nerve cords [32]. The extracellular (field) potential generated by the lateral giant spike is large enough to be recorded outside the animals during a naturally evoked tail-flip [33]. These large extracellular fields make the lateral giant neurons ideal biological models to be interfaced with STNOs and produce a microwave signal.

**Materials and methods**

The STNO that we employ in this work is an elliptical magnetic tunnel junction nanopillar with lateral dimensions 50 nm × 190 nm. Fig. 1.a shows the complete layer structure for the device, with thicknesses (in nanometers) indicated in parentheses. We deposited all layers using magnetron sputtering in a Singulus TIMARIS system, and patterned the magnetic tunnel junctions using electron beam lithography followed by ion milling. The synthetic antiferromagnet is $PtMn/Co_{70}Fe_{30}/Ru/Co_{40}Fe_{40}B_{20}$ with the $Co_{70}Fe_{30}$ pinned layer and the $Co_{40}Fe_{40}B_{20}$ reference layer antiferromagnetically coupled by the tuned thickness of Ru. Prior to patterning, we anneal the multilayer for 2 hours at 300 °C in a 1 T in-plane field to set the pinned layer exchange bias direction parallel to the long axis of the nanopillars.

We obtained adult crayfish (*Procambarus clarkii*) of both sexes from a commercial supplier and kept them in large communal tanks before the experiments. Individual animals (total body lengths 7-10 cm, measured from rostrum to telson) were anaesthetized on ice for several minutes until immobility. We separated the abdomen from the anterior part of the body and pinned it down in a petri dish. We removed the membrane covering the ventral nerve cord and muscles, cut all ganglionic nerves, and dissected out the ventral nerve cord. Next, we firmly pinned down the ventral nerve cord dorsal side up in a round petri dish lined with silicone elastomer (Sylgard) and filled with fresh crayfish saline (Fig. 1.b). The saline in the dish maintained a constant temperature of 20-21°C throughout the experiments. Only preparations that appeared healthy were used, which allowed continuous measurements for several hours after the dissection.

We placed a pair of silver wire electrodes on the upper side surface of the ventral nerve cord to stimulate the lateral giant neuron and a second identical pair of electrodes near the frontal end of the nerve cord to record the lateral giant action potential (Fig. 1.b). To evoke lateral giant action potentials, we applied voltage pulses with amplitudes of 5-10 V and pulse durations of 0.2-0.5 ms to the ventral nerve cord. We stimulated using a data acquisition board (NI USB-6211) controlled by a LabVIEW (National Instruments) program. We used a differential amplifier (A-M Systems, Model 1700) for stimulations and recordings. A stimulus isolation unit (Grass, Model SIU5) applied a constant voltage stimulus. We used an amplifier with 1000x of gain to amplify the recorded signals, and then measured the signal using an oscilloscope (Lecroy HRO 64Zi).

To drive the STNO with extracted signal from the crayfish, we utilized the experimental configuration shown in Fig. 1.c. We placed the STNO in a home-built probe station and connected to the input and output leads using a non-magnetic picoprobe (10-50/30-125-BeCu-2-R-200, GGB industries). An electromagnet applied a magnetic field along the in-plane minor axis of the device to produce precession of the magnetic free layer [34]. We connected the silver electrodes from the crayfish neuron to the input port of the device. A bias tee separated out the direct electrical signal from the neuron from the induced microwave field oscillations in the STNO. This technique provides access to both signals and enables us to compare the direct neuron activity to the device microwave response. We measured the electrical action potential from the lateral giant neuron using an oscilloscope (Lecroy HRO 64Zi). We measured the microwave signal using a low noise amplifier (Pasternack PE15A1013) and a spectrum analyzer (Agilent 8564 EC). In order to improve signal-to-noise, we averaged the output microwave signal over 1000-6000 stimulation pulses at frequencies between 0.5-20 Hz. The lateral giant action potential can be easily identified by its characteristic shape, large amplitude, fast conduction velocity, and low firing threshold. However, depending on electrode placement, recordings can vary across preparations. This variability was more prominent when using the STNO because it required using one recording electrode as a ground electrode; among some other minor effects, this substantially reduced the amplitude of the recorded action potential. We stimulated the crayfish neuron at subthreshold level as controls to confirm that recording experiments evoked neural activity. Since thousands of stimuli at high frequency can lead to occasional lateral giant spike failure, we obtained best results with longer inter-stimulus intervals. In total, we were able to successfully record three examples of neural activity in the isolated ventral nerve cord from three different preparations using the STNO.

**Results and discussion**

We first characterized the STNO properties to determinate the strength of the external magnetic field that produced the maximum output. We drove the device with an external power supply and monitored the microwave response. Fig. 2 shows an average of 100 acquisitions of the power spectral density of the STNO output for different direct voltages as well as the measurements for the optimal external magnetic field of 10 mT that produced the maximum microwave power output. These measurements show the device oscillates with a frequency of 1.2 GHz and with amplitudes ranging from 0.05 fW/MHz to 0.42 fW/MHz for the given voltages from 0.25 mV to 0.75 mV, respectively. From the measured resistance of the device (600 Ω), the voltage range corresponds to a peak input current in the range of 0.4 µA to 1.25 µA. Based on these currents, we conclude that the device works in the sub-threshold regime where the applied current is below the critical current for zero-temperature onset of self-oscillations. In this case, the observed microwave signal arises from temperature-induced precession of magnetization of the free layer [35].

Using the optimized magnetic field, we applied repeated electrical stimuli to the lateral giant neuron to evoke action potentials. Fig. 3.a shows the direct electrical signal from a crayfish neuron that we extracted from the inductive port of the bias tee. The black trace shows the voltage that we measured with the oscilloscope at the recording electrodes. In this specific case, the stimulus was a square pulse with an amplitude of 10 V and a duration of 0.2 ms. The initial spike (before 0.5 ms) in the voltage trace is the stimulus artifact due to direct coupling of the electrical signal from the stimulus electrode to the recording electrode. At approximately 1 ms after the stimulus, we observe a second voltage pulse that corresponds to an action potential, which reaches a maximum amplitude of 0.23 mV.

We next drove the STNO device directly with the output voltage of the neuron. The red trace in Fig. 3.a shows the microwave power versus time at the frequency of 1.2 GHz with a bandwidth of 2 MHz. To increase the signal-to-noise ratio we recorded an average of 3,000 sweeps (i.e., 3000 stimulations of the neuron). The instantaneous microwave output power follows the voltage waveform of the action potential. The large voltage at the beginning is due to direct activation of the STNO by the stimulus artifact. Following the stimulus artifact, we recorded another peak power 1 ms later, which matches the action potential. We found that the peak power that coincides with the action potential had a magnitude of 0.08 fW.

To determine the sensitivity of the device and assess the strength of the input signal, we generated artificial action potentials using an analog voltage waveform generator. We stimulated the STNO with input voltage pulses with temporal waveforms identical to those produced by a crayfish neuron (Fig. 3.b inbox blue line). We artificially generated the action potential with the data acquisition board and the Labview software and inputted it directly to the STNO by the inductive port of the bias tee. We employed action potentials with amplitudes of 0.25 mV, 0.6 mV and 1.2 mV. Fig. 3.b shows the microwave power output from the STNO versus time for artificial action potentials with a peak voltage range from 0.25 mV to 1.2 mV at a frequency of 1.2 GHz, with a bandwidth of 2 MHz. To discriminate the signal from the noise, we needed to average 1,000 acquisitions for the 0.25 mV action potential, 100 acquisitions for the 0.6 mV and 10 acquisitions for 1.2 mV. We observed peak power of 0.1 fW, 0.8 fW and 2.8 fW for action potential amplitudes of 0.25 mV, 0.6 mV and 1.2 mV, respectively. The power that we obtained for 0.25 mV with the artificial action potential agrees well with the result shown in Fig. 3.a. This result shows that increasing the peak voltage by 4.8 times produced a 28-fold increase in the peak power making the detection easier. Taking into account the device resistance (600 Ω) input currents on the order of microamps could provide single action potential detection.

## Conclusion

In summary, we demonstrated that action potentials from crayfish isolated nerve cords can generate microwave signals in STNOs. The current device produced a peak power of 0.08 fW, which required averaging 3,000 action potentials from a single neuron. A number of approaches could significantly increase this power to the regime, which would allow us to measure a single action potential. Devices with lower threshold currents could significantly increase the output power [36]. Devices with large-amplitude magnetization precession [34], [37], reduced phase noise [38], oscillators with large volume of the free magnetic layer [39] or Phase locked oscillators [40], [41] could further improve the sensing by emitting more power in a narrower bandwidth. Another possibility is to stimulate simultaneously many spintronic devices attached to a single neuron. According to the microwave output detected in the present work, an array of 100 oscillators, fitting in an area over 1 $\mu m^2$, could produce sufficient microwave power to detect single action potentials. Improved electrodes with reduced contact resistance could also potentially increase the input current to the STNO, thereby significantly increasing the generated microwave signal. Alternatively, glucose fuel cells can bias an amplifier able to increase the input current to the STNO [42].

Ultimately, our results open up a new approach for high resolution sensing of bioelectric signals using spintronic devices. STNOs occupy a small device footprint, potentially in the nanoscale, and operate at low input currents, opening up the possibility for extremely dense low-power wireless sensor arrays. Furthermore, the oscillation frequency of these devices is highly tunable through the external magnetic field [22], [26]. In the presence of a strong magnetic field gradient, this property could enable STNOs to encode their position in the oscillation frequency in an analogous way to magnetic resonance imaging. Furthermore, the small size of these devices opens up the possibility to introduce them intravenously. Previous studies showed that magnetic particles of similar dimensions can cross the blood-brain-barrier and reach targets in the brain without disrupting the barrier in rat models[43], [44]. In addition to neuronal sensing, spintronic sensors could be useful for detecting electrical signals from other tissue such as heart, or other muscles. These properties could significantly enhance and extend current biological sensing capabilities.

## ACKNOWLEDGEMENT

We thank Dr. John Rodger and Bisrat Adissie for providing access to the microwave equipment. We also thank Pablo Villar del Rio who provided comments on this work. This work was supported by a seed grant from the Brain and Behavior Initiative (BBI) at the University of Maryland, College Park. We gratefully acknowledge support from a NSF BRAIN EAGER grant (grant number DBI1450921) as part of the BRAIN initiative. The work of Yu-Jin Chen and Ilya Krivorotov on sample design and characterization was supported as part of the SHINES, and Energy Frontier Research Center funded by the U.S. Department of Energy, Office of Science, Basic Energy sciences under Award # SC0012670.

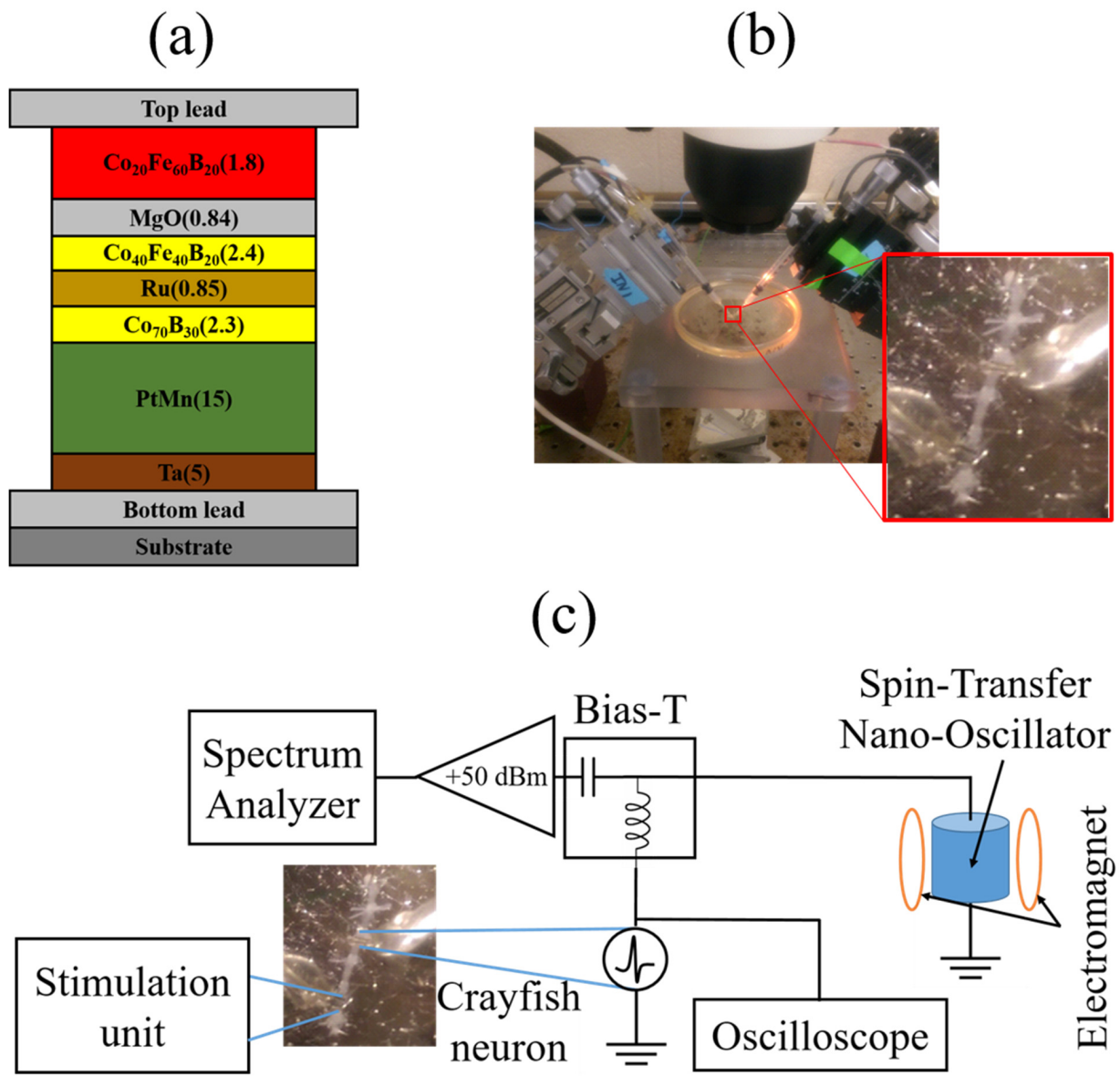

Figure 1. (a) Schematic of the nanopillar spin-torque nano-oscillator device. The numbers in parentheses are the layer thicknesses in nanometers (b) Picture of experimental setup used for crayfish neuron stimulation and recording. (c) A schematic of the circuit to trigger the spin-torque nano-oscillators with action potentials from crayfish neuron.

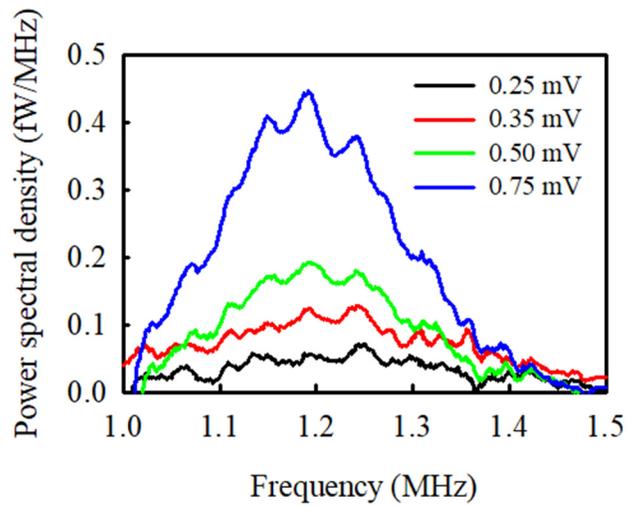

Figure 2. The power spectral density of the microwave signal measured from the spin-torque nano-oscillator for different input direct voltages with an in-plane magnetic field of 10 mT along the minor axis.

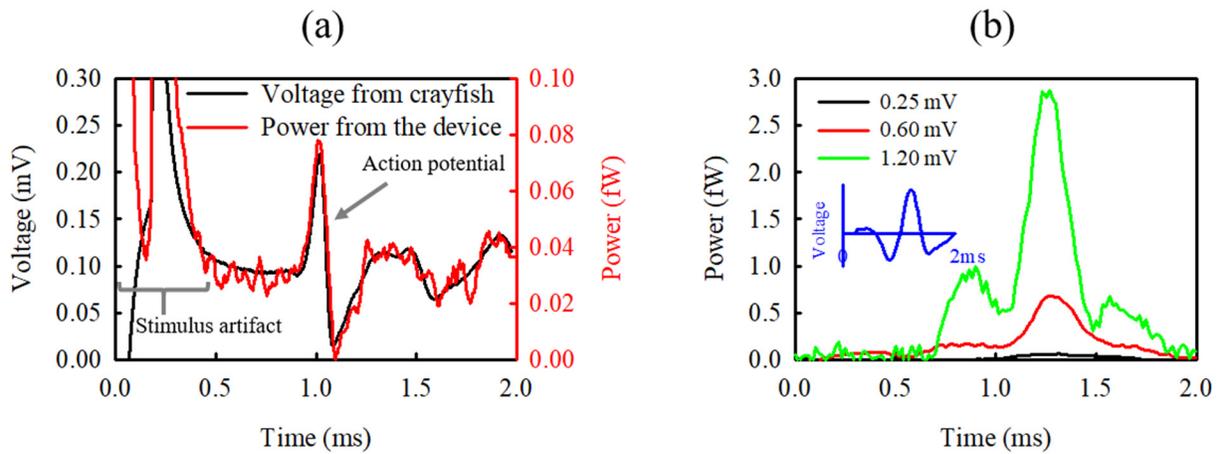

Figure 3. (a) Action potential recorded from crayfish neurons using silver electrodes (black) and the corresponding microwave power measured from the spin-torque nano-oscillator (red). (b) Microwave power from the spin-torque nano-oscillator when excited with an artificial action potential (inbox blue curve) of different amplitudes.